\documentclass[
reprint,
superscriptaddress,
amsmath,
amssymb,
aps,
longbibliography,
pra,
showpacs,
floatfix
]{revtex4-1}
\usepackage{hyperref}
\usepackage{graphicx}
\usepackage{xcolor} 
\usepackage{booktabs}

\usepackage{tabularx}
\usepackage{multirow}

\usepackage{braket}

\usepackage{comment}
\usepackage{enumitem}
\usepackage{soul}

\definecolor{color1}{rgb}{0,0.25,0.70}
\hypersetup{colorlinks=true,
    linkcolor={color1},
    citecolor={color1},
    urlcolor={color1}
}

\begin{document}

\title{Dynamically induced multiferroic polarization}

\author{Carolina Paiva}
\email{paiva@mail.tau.ac.il}
\affiliation{School of Physics and Astronomy, Tel Aviv University, Tel Aviv 6997801, Israel}

\author{Michael Fechner}
\affiliation{Max Planck Institute for the Structure and Dynamics of Matter, Center for Free-Electron Laser Science (CFEL), 22761 Hamburg, Germany}

\author{Dominik~M.\ Juraschek}
\email{djuraschek@tauex.tau.ac.il}
\affiliation{School of Physics and Astronomy, Tel Aviv University, Tel Aviv 6997801, Israel}

\date{\today}


\begin{abstract}
We describe a mechanism by which both ferroelectric polarization and magnetization can be created in nonpolar, nonmagnetic materials. Using a combination of phenomenological modeling and first-principles calculations, we demonstrate that ferroelectric polarization, magnetization, or both simultaneously can be transiently induced by an ultrashort laser pulse upon linearly, circularly, or elliptically polarized excitation of phonon modes in $\gamma$-LiBO$_2$. The direction and magnitude of the multiferroic polarization can be controlled by the chirality of the laser pulse and the phonon modes, offering a pathway for controlling multiferroicity and magnetoelectricity on ultrafast timescales. 
\end{abstract}

\maketitle


\section{Introduction}

Light-induced control of magnetism and ferroelectricity promises applications in future data processing and storage devices that operate on ultrafast timescales of femto- and picoseconds, orders of magnitude faster than state-of-the-art technology \cite{Nemec2018,Kimel2020,Disa2021}. A key challenge lies in finding new mechanisms of switching and generating ferroic polarizations. A particularly intriguing way of controlling ferroic order involves driving lattice vibrations into the nonlinear regime with ultrashort laser pulses, which modifies the geometry of the crystal lattice and therefore the electronic correlations in the solid. In recent years, a broad variety of theoretical and experimental studies have demonstrated controlling and inducing of ferroelectricity \cite{Qi2009,Katayama2012,subedi:2015,Chen2016,subedi:2017,Mankowsky_2:2017,Li2019,Nova2019,Mertelj2019,Park2019,Shin2020,Abalmasov2020,Pal2021,Abalmasov2021,Kong2021,Henstridge2022,Shin2022,Chen2022,Zhang2022,BustamanteLopez2023,Cheng2023,Li2023,Zhuang2023,Kwaaitaal2023,Zhilyaev2023,Zhilyaev2024,Khalsa2024} and magnetism \cite{Fechner2018,Khalsa2018,Gu2018,Radaelli2018,Rodriguez-Vega2020,Disa2020,Afanasiev2021,Stupakiewicz2021,Mashkovich2021,Mertens2021,Formisano2022,Rodriguez-Vega2022,Padmanabhan2022,Disa2023,Giorgianni2023,nova:2017,juraschek2:2017,Shin2018,Juraschek2019,Juraschek2020_3,Geilhufe2021,Juraschek2022_giantphonomag,Geilhufe2023,Basini2024,Davies2024,Romao2024_NV,Luo2023,Nielson2023,Gao2023} through coherent and nonlinear phonon driving. Despite these successes, the simultaneous generation of multiple ferroic orders on demand has yet remained elusive.

\begin{figure}[b]
\centering
\includegraphics[width=1\linewidth]{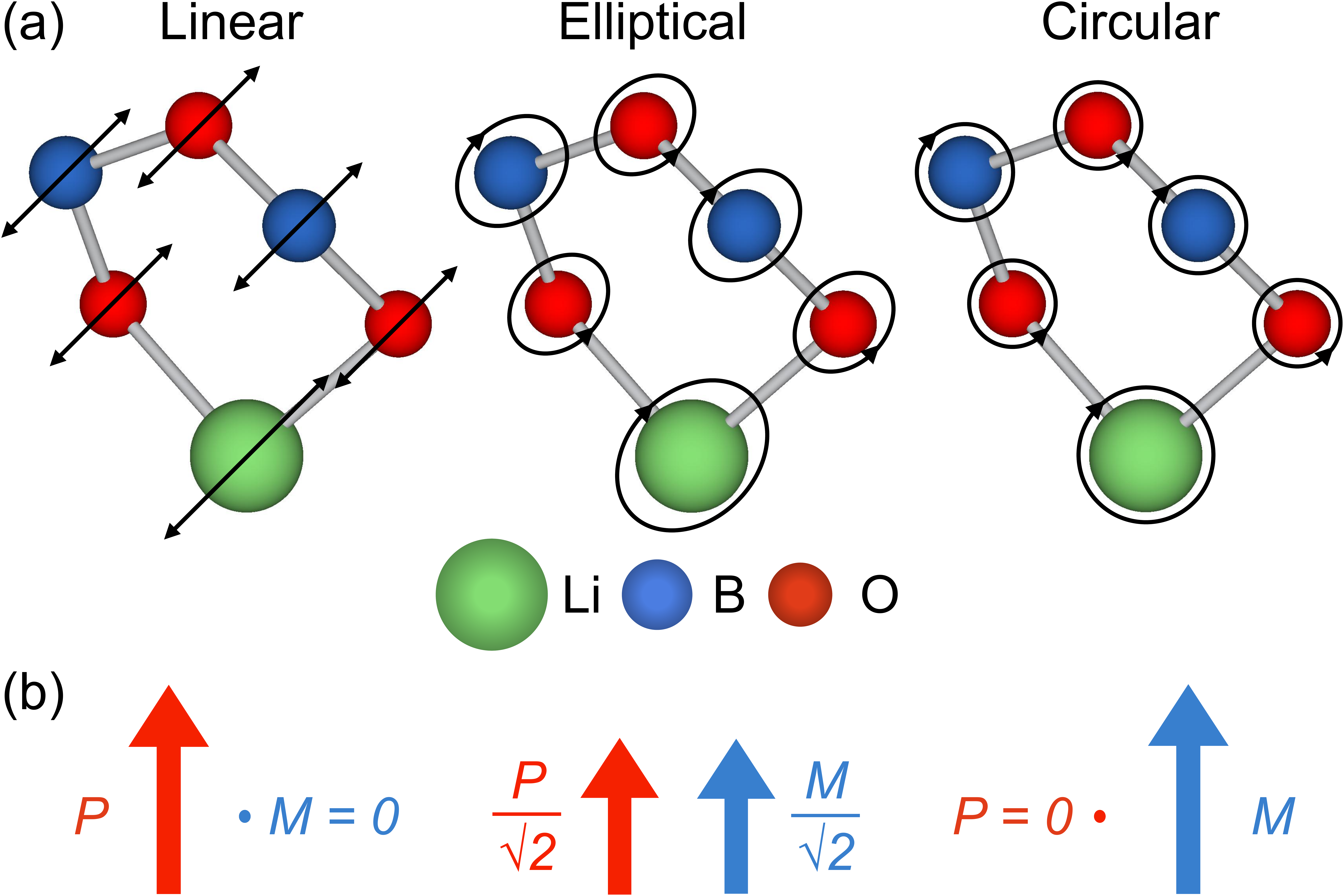}
\caption{
Chirality-dependent ferroic polarizations. 
(a) Ionic motions along the eigenvectors of a linearly, elliptically, and circularly excited $E$ mode in one of the borate rings of $\gamma$-LiBO$_2$.
(b) Linear excitation induces only a ferroelectric polarization, $P$, elliptical excitation induces both a ferroelectric polarization and a magnetization, $M$, and circular excitation induces only a magnetization.
}
\label{fig:Conceptual_Picture}
\end{figure}

Here, we theoretically demonstrate the transient generation of both ferroelectric polarization, $P$, and magnetization, $M$, in a nonpolar, nonmagnetic material through nonlinear phonon excitation by an ultrashort mid-infrared (mid-IR) pulse. We show that the chirality of the pulse and therefore the coherently excited phonon modes can be used to selectively generate each polarization individually or both simultaneously, see Fig.~\ref{fig:Conceptual_Picture}. The directions of $P$ and $M$ can be switched by changing the orientation and handedness of the laser pulse. We use phenomenological modeling and first-principles calculations to simulate the polarization and magnetization dynamics in $\gamma$-LiBO$_2$, which we show can be described by second-order nonlinear electric and magnetoelectric responses to the incident laser pulse. The  mechanism can be measured in ultrafast pump-probe experiments and provides a route towards controlling multiferroicity and magnetoelectricity on ultrafast timescales.


\section{Theory}


\subsection{Properties of LiBO$_2$}
%
$\gamma$-LiBO$_2$ is a nonpolar, nonmagnetic material, crystallizing in the noncentrosymmetric tetragonal space group $I\bar{4}2d$. Its primitive unit cell consists of 8 atoms and hosts 24 phonon modes, characterized by the irreducible representations $A_1$, $A_2$, $B_1$, $B_2$, and $E$. The degenerate IR-active $E$ modes are polarized in the $ab$-plane of the crystal, whereas the $B_2$ modes are polarized along the $c$-axis. $\gamma$-LiBO$_2$ has a band gap of 10~eV ~\cite{Basalaev2019}, making it an excellent candidate for phonon pumping in the mid-IR, far from electronic resonances. A previous group-theory analysis by Radaelli suggested that the point group of $\gamma$-LiBO$_2$ enables light-induced ferroelectricity, connected to its piezoelectric properties \cite{Radaelli2018}, see Appendix. This symmetry allows for a ferroelectric polarization along the $c$-axis of the crystal upon displacements of the atoms along the eigenvectors of the $B_2$ modes, which changes the point group from $\bar{4}2 m$ to the polar $mm2$ group. We combine this feature with light-induced magnetization following the coherent excitation of chiral phonon modes \cite{juraschek2:2017,Juraschek2019,Geilhufe2021} to generate multiferroic polarization.


\subsection{Nonlinear phonon dynamics}
%
We model the coherent nonlinear phonon dynamics of the $E$ and $B_2$ modes with the equations of motion \cite{subedi:2014,fechner:2016,Juraschek2018}

\begin{equation}
    \ddot{Q}_\alpha + \kappa_\alpha \dot{Q}_\alpha + \partial_{Q_\alpha} V  = \mathbf{Z}_{\alpha} \cdot \mathbf{E}(t), \label{eq:phononeom}
\end{equation}
where $Q_\alpha$ is the amplitude of phonon mode $\alpha\in\{a,b,c\}$, $\kappa_\alpha$ is its linewidth, and $V$ is the phonon potential energy. $\mathbf{Z}_{\alpha}=\sum_n Z_{n}^* \mathbf{q}_{n,\alpha}/\sqrt{M_n}$ is the mode effective charge, where $Z_{n}^*$ is the Born effective charge tensor, $\mathbf{q}_{n,\alpha}$ the phonon eigenvector, and $M_n$ the atomic mass of atom $n$, and the sum runs over all atoms in the unit cell. $\mathbf{E}$ is the electric field component of the laser pulse in the material. The anharmonic phonon potential can be written as

\begin{equation}
\label{eq:phononpotential}
    V  =  \frac{\Omega_{0}^2}{2}Q_{a}^2+\frac{\Omega_{0}^2}{2}Q_{b}^2+\frac{\Omega_{c}^2}{2}Q_{c}^2 + c Q_{a}Q_{b}Q_{c} + \tilde{V}.
\end{equation}
where $\Omega_a=\Omega_b\equiv\Omega_0$ is the eigenfrequency of the $E$ mode and $\Omega_c$ that of the $B_2$ mode. The coefficient $c$ denotes the trilinear phonon coupling primarily of interest here. In $\tilde{V}= d_a Q_a^4 + d_b Q_b^4 + d_c Q_c^4 + d_{ab} Q_a^2 Q_b^2 + d_{ac} Q_a^2 Q_c^2 + d_{bc} Q_b^2 Q_c^2 + g_a Q_a^6 + g_b Q_b^6 + g_c Q_c^6 $, we include single-mode anharmonicities up to sixth order and nonlinear phonon couplings up to quartic order, denoted by $d_\alpha$, $d_{\alpha\beta}$, and $g_\alpha$ respectively. Cubic- and fifth-order anharmonicities, $Q_\alpha^3$ and $Q_\alpha^5$, and quadratic-linear couplings, $Q_{\alpha}^2 Q_\beta$, are symmetry-forbidden, $\alpha\neq\beta\in\{a,b,c\}$.

The electric field component of the laser pulse lies in the $ab$-plane, $\mathbf{E}(t)=\mathcal{E}(t)(\cos(\omega_0 t),\cos(\omega_0 t + \phi), 0)$. $\mathcal{E}(t) = \epsilon_\infty^{-1} E_{0} \exp[-t^2/(\tau\sqrt{8\ln 2})^2]$ is the carrier envelope, $\epsilon_\infty$ is the optical dielectric constant of the material, $E_{0}$ is the peak electric field in free space, $\tau$ is the full width at half maximum pulse duration, and $\omega_{0}$ is the center frequency. $\phi$ is the carrier envelope phase determining the polarization (linear, elliptical, circular) of the pulse. The equations of motion then read

\begin{align}
\label{eq:Q_1_time}
 \ddot{Q}_{a} + \kappa_{0}\Dot{Q}_{a} + \Omega^2_{0}Q_{a} & = Z_{a,x}E_{x}(t) - c Q_{b} Q_{c} - \partial_{Q_a} \tilde{V}, \\
\label{eq:Q_2_time}
 \ddot{Q}_{b} + \kappa_{0}\Dot{Q}_{b} + \Omega^2_{0}Q_{b} & = Z_{b,y}E_{y}(t) - c Q_{a} Q_{c} - \partial_{Q_b} \tilde{V},\\
\label{eq:Q_c}
 \ddot{Q}_{c} + \kappa_{c}\Dot{Q}_{c} + \Omega^2_{c}Q_{c} & = -cQ_{a}Q_{b} - \partial_{Q_c} \tilde{V},
\end{align}
where we set $\kappa_a=\kappa_b\equiv\kappa_0$. The main driving force for the two components of the $E$ mode is the electric field component of the pulse. For the $B_2$ mode, the driving force is proportional to $Q_a Q_b$. The back-action of the $B_2$ mode on the $E$ modes, contained in the $Q_{a/b} Q_{c}$ and in the higher-order terms in the equations of motion is much weaker than the driving force of the laser pulse itself.

Solving the equations in the frequency domain yields analytical expressions to first order in the electric field for $Q_{a}$ and $Q_{b}$ and to second order for $Q_{c}$ \cite{kahana2023lightinduced,BustamanteLopez2024}, 
\begin{align}
\label{eq:Q_1/2_resultfrequency}
Q_{a}(\omega) & = Z_{a,x}\frac{E_{x}(\omega)}{\Delta_{0}(\omega)}, ~ Q_{b}(\omega) = Z_{b,y}\frac{E_{y}(\omega)}{\Delta_{0}(\omega)},\\
Q_c(\omega) & = -\frac{cZ_{a,x} Z_{b,y}}{\Delta_{c}(\omega)}\left(\frac{E_x(\omega)}{\Delta_{0}(\omega)}\circledast \frac{E_y(\omega)}{\Delta_{0}(\omega)}\right),
\label{eq:Q_c_resultfrequency}
\end{align}
Here, $\Delta_{\alpha} (\omega) = \Omega^{2}_\alpha-\omega^2+i\omega \kappa_{\alpha}$ and $\circledast$ denotes the convolution operator. Please see Appendix for details of the derivations. In the following, we will show how the nonlinear phonon dynamics lead to dynamically induced ferroelectric polarization and magnetization.


\subsection{Ferroelectric polarization from phononic rectification}
%
The central component for light-induced ferroelectricity is the trilinear coupling term $Q_{a}Q_{b}Q_{c}$ in the phonon potential energy, Eq.~\eqref{eq:phononpotential}. The IR-active $B_2$ mode produces a polarization along the $c$-axis of the crystal, $P_{c,z}=Z_{c,z} Q_c/V_c$, where $V_c$ is the unit-cell volume. A mere oscillatory excitation of the $B_2$ mode, for example through impulsive stimulated or ionic Raman scattering \cite{Hortensius2020,Neugebauer2021}, would lead to an oscillating polarization. Therefore, the driving force acting on the $B_2$ mode in Eq.~\eqref{eq:Q_c} needs a unidirectional component, which causes the mean of the $B_2$ mode to follow the mean square of the $E$ mode components, $\langle Q_c \rangle \sim \langle Q_a Q_b \rangle \neq 0$  \cite{subedi:2014,mankowsky:2014,juraschek:2017}. This is known as phononic rectification and in the case of $\gamma$-LiBO$_2$ can be achieved with linearly polarized excitation of the $E$ mode oriented at 45$^\circ$ in the $ab$-plane of the crystal ($\phi=0$). 

For example, a simple linearly polarized oscillation, $Q_a = Q_b = \sin(t)$, yields $\langle Q_a Q_b \rangle = 0.5$, causing maximum rectification and therefore unidirectional displacement of the $B_2$ mode. Reorienting the polarization of the laser pulse by 90$^\circ$ in the $ab$-plane (i.e. setting $\phi=\pi$) changes the sign of the force and therefore reverts the direction of rectification \cite{juraschek:2017,Radaelli2018}. In contrast, a circularly polarized excitation ($\phi=\pi/2$) produces no net force and therefore no rectification, as $Q_a = \sin(t)$ and $Q_b = \sin(t+\pi/2)$ yields $\langle Q_a Q_b \rangle = 0$. Finally, an elliptically polarized excitation ($\phi=\pi/4$) produces a reduced rectification with respect to the linear case, as $Q_a = \sin(t)$ and $Q_b = \sin(t+\pi/4)$ yields $\langle Q_a Q_b \rangle = 0.5/\sqrt{2}$. The dependence of the ferroelectric polarization on the chirality of the excitation is shown in Fig.~\ref{fig:Conceptual_Picture}.

Having obtained analytical expressions for the phonon modes driven by the electric field component of the laser pulse in Eqs.~\eqref{eq:Q_1/2_resultfrequency} and \eqref{eq:Q_c_resultfrequency}, we now turn our attention to deriving the response of the ferroelectric polarization. Substituting these results into $P_{c,z}(\omega)=Z_{c,z} Q_c(\omega)/V_c$ and assuming monochromatic light ($\tau\rightarrow\infty$) we obtain a second-order nonlinear polarization given by

\begin{equation}
P_{c,z}(0) = \epsilon_{0}\chi^{(2)}_{e,xyz}(0;\omega_0,-\omega_0) E_{x}(\omega_0)E_{y}^*(\omega_0).
\end{equation}
Here, $\epsilon_{0}$ is the vacuum permittivity and $\chi^{(2)}_{e,xyz}=\chi^{(2)}_{e,yxz}$ is a second-order nonlinear electric susceptibility induced by the nonlinear phonon coupling, given by

\begin{equation}
\chi^{(2)}_{e,xyz}(0;\omega_0,-\omega_0) = \frac{-c}{\sqrt{2\pi}\epsilon_{0}V_c} \frac{Z_{a,i} Z_{b,j} Z_{c,k}}{\Delta_{c}(0)|\Delta_{0}(\omega_0)|^2}.
\label{eq:nonlinear_electric_susceptibility}
\end{equation}
$\chi^{(2)}_{e}$ characterizes the static (ferroelectric) polarization generated by the $B_2$ mode when the two components of the $E$ mode are driven by a laser with frequency $\omega_0$. For the full frequency-dependent response, see Appendix.


\subsection{Magnetization from chiral phonons}
%
In addition to the ferroelectricity arising from phononic rectification, an elliptically or circularly polarized mid-IR pulse can induce a magnetization through the excitation of a chiral phonon mode. The elliptical or circular superposition of the two components of the $E$ mode in the $ab$-plane produces orbital motions of the ions around their equilibrium positions, generating a magnetization perpendicular to the plane \cite{nova:2017,juraschek2:2017,Shin2018,Juraschek2019,Juraschek2020_3,Geilhufe2021,Juraschek2022_giantphonomag,Xiong2022,Geilhufe2023,Basini2024,Davies2024,Romao2024_NV,Luo2023,Nielson2023,Zhang2023chiral,Gao2023,kahana2023lightinduced,Chaudhary2023}. The magnetization is given by $M_z = \mu_{ph} L_z/(\hbar V_c)$, where $\mu_{ph}$ is the phonon magnetic moment and $\mathbf{L} = \mathbf{Q}\times \dot{\mathbf{Q}}$ is the phonon angular momentum with $\mathbf{Q} = (Q_{a},Q_{b},0)$. In the ionic point-charge picture, the phonon magnetic moment is given by $\mu_{ph} = \hbar^{-1}\sum_n Z_n^*/(2M_n)\mathbf{q}_{n,a}\times\mathbf{q}_{n,b}$. If we set $Q_a = \sin(t)$ and $Q_b = \sin(t+\phi)$, the magnetization becomes $M_z=\sin(\phi)$. The magnetization is therefore maximal for a circularly polarized excitation ($\phi=\pi/2$), $M_z=1$, reduced for an elliptically polarized excitation ($\phi=\pi/4$), $M_z=1/\sqrt{2}$, and vanishing for a linearly polarized excitation ($\phi=0$), $M_z=0$. The dependence of the magnetization on the chirality of the excitation is illustrated in Fig.~\ref{fig:Conceptual_Picture}.

Substituting Eq.~\eqref{eq:Q_1/2_resultfrequency} into $M_z(\omega)=\mu_{ph} L_z(\omega)/(\hbar V_c)$, we obtain a second-order nonlinear magnetization,

\begin{equation}
    M_{z}(0) = \chi^{(2)}_{me,ijk}(0;\omega_0,-\omega_0)E_{i}(\omega_0)E_{j}^*(\omega_0),
\end{equation}
where $\chi^{(2)}_{me,ijz}=-\chi^{(2)}_{me,jiz}$ is a second-order nonlinear magnetoelectric susceptibility induced by the phonon magnetic moment and given by

\begin{equation}
    \chi^{(2)}_{me,ijk}(0;\omega_0,-\omega_0) = \epsilon_{ijk} \frac{\mu_{ph}}{\sqrt{2\pi} \hbar V_c}\frac{i\omega_0 Z_{a,i}Z_{b,j}}{|\Delta_{0}(\omega_0)|^2},
    \label{eq:magnetoelectric_susceptibility}
\end{equation}
where $\epsilon_{ijk}$ is the Levi-Civita tensor. $\chi^{(2)}_{me}$ characterizes the static magnetization generated by the $E$ modes when driven by a laser with frequency $\omega_0$. For the full frequency-dependent response, see Appendix.


\section{Numerical evaluation}


\subsection{First-principles calculations}
%
We calculate the phonon eigenfrequencies, eigenvectors and Born effective charges of $\gamma$-LiBO$_2$ using the density functional theory formalism as implemented in \textsc{vasp} \cite{kresse:1996,kresse2:1996}, as well as the frozen-phonon method as implemented in \textsc{phonopy} \cite{phonopy}. We used the default PAW pseudopotentials with valence electron configurations Li($2s^1$), B($2s^2 2p^1$), and O$(2s^2 2p^4)$. We used the PBEsol form of the generalized gradient approximation (GGA) for the exchange-correlation functional \cite{perdew:2008}. No Hubbard or Hund's exchange correction was added. We converged the Hellmann-Feynman forces to $10^{-5}$~eV/\AA{} using a plane-wave energy cutoff of 600~eV and a $12\times12\times12$ k-point mesh. The lattice constants of our fully relaxed structure, $a=4.19$~\AA{} and $c=6.53$~\AA{} ($V_c=57$~\AA$^3$), fit reasonably well to experimental values \cite{Marezio1965} and to previous calculations \cite{Basalaev2019}. To obtain the single-mode anharmonicities and nonlinear phonon couplings, we computed the total energy of the system on a $11\times11\times11$ grid of atomic displacements along the eigenvectors of each of the phonon modes, parametrized by the phonon amplitudes $Q_a$, $Q_b$, and $Q_c$, and fitted the resulting potential energy landscape to the phonon potential energy, $V$.

We show the calculated phonon parameters in Table~\ref{table:anharmonic_coefficients}. Phonon frequencies of all IR-active modes are shown in the Appendix.  The coefficients $c$ and $d_{ac}=d_{bc}$ for the $B_2$(25.7) mode are $-2.2$~eV/(\AA$\sqrt{u}$)$^3$ and 1.3~eV/(\AA$\sqrt{u}$)$^4$, respectively. To our knowledge, these are the largest nonlinear phonon couplings reported in literature, more than three times larger than the leading coefficients in YBCO 
\cite{fechner:2016}. The product of the trilinear phonon couplings and mode effective charges, $cZ_{c,z}$, has the same sign for all three $B_2$ modes in the system and therefore leads to a cooperative ferroelectric polarization response according to Eq.~\eqref{eq:nonlinear_electric_susceptibility}. We further obtain an optical dielectric constant in the $ab$-plane of $\epsilon_\infty = 2.9$ and a phonon magnetic moment of $\mu_{ph} = 0.14~\mu_N$, where $\mu_N$ is the nuclear magneton. For the phonon linewidths, we assume phenomenological values of $\kappa_\alpha = 0.1\Omega_\alpha/(2\pi)$.


\begin{table}[t]
\caption{
Calculated eigenfrequencies in THz, single-mode anharmonicities and nonlinear phonon couplings in meV/(\AA{}$\sqrt{u})^n$, $n$ being the order of the phonon amplitude and $u$ the atomic mass unit, and mode effective charges in $e/\sqrt{u}$, where $e$ is the elementary charge. 
}
\label{table:anharmonic_coefficients}
\begin{tabular}{llllllllllll}
\hline
 & $c$ & $d_a$ & $d_b$ & $d_c$ & $d_{ab}$ & $d_{ac}$ & $d_{bc}$ & $g_a$ & $g_b$ & $g_c$ & $Z_{c,z}$ \\ 
\hline
$E(26.2)$ & & 211 & 211 & & -4 & & & 22 & 22 & & 1 \\
$B_{2}(9.05)$ & 179 & & & -13 & & 48 & 48 & & & 0.7 & 0.7\\
$B_{2}(19.8)$ & 772 & & & 3 & & -63 &-63 & & & -0.3 & 0.5 \\
$B_{2}(25.7)$ & -2157 & & & 11 & & 1315 &1315 & & & -0.2 & -1.2\\
\hline
\end{tabular}
\end{table}


\begin{figure*}[t]
\centering
\includegraphics[width=0.85\textwidth]{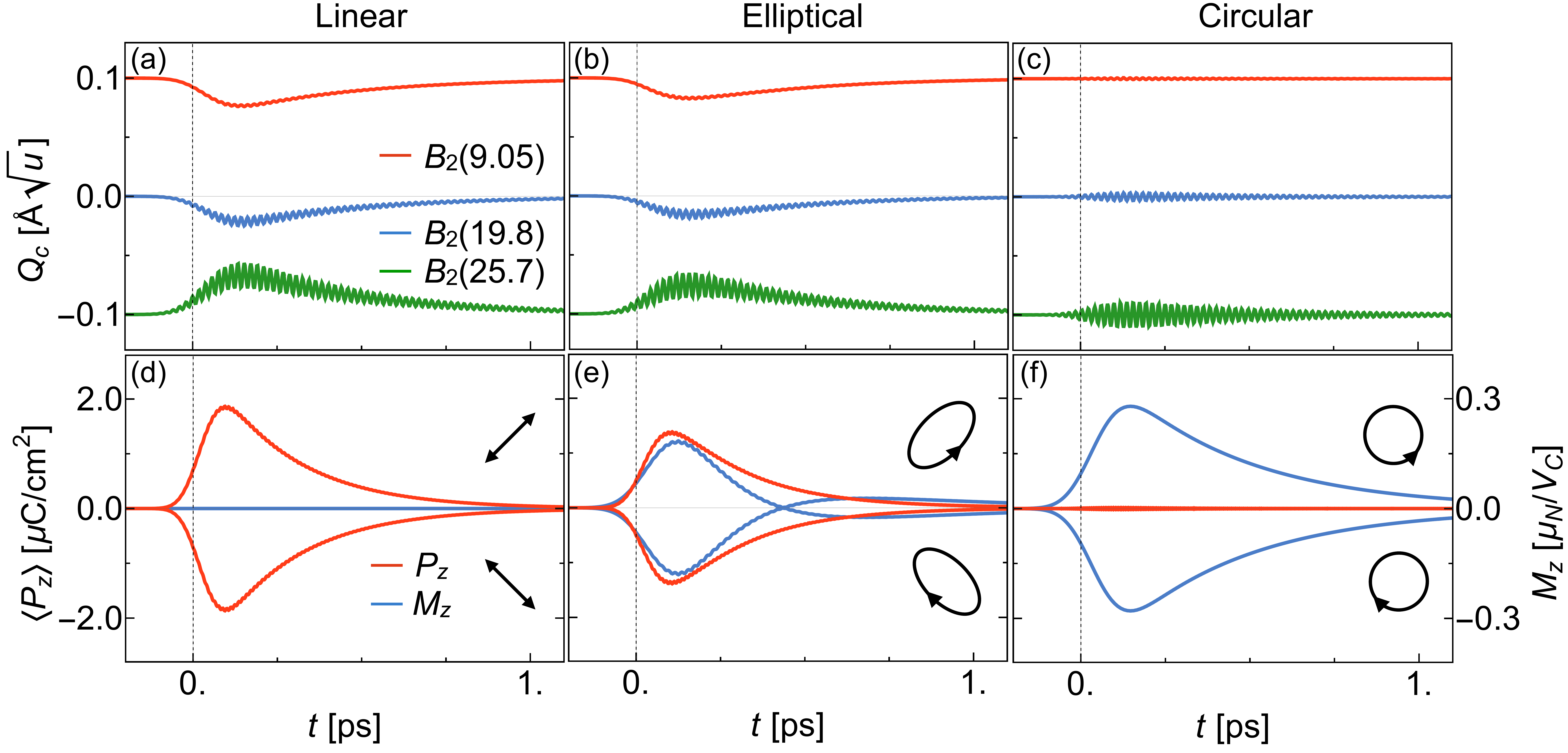}
\caption{
{\small
Ferroelectric polarization and magnetization dynamics in response to the resonant excitation of the $E$ mode at 26.2~THz with linearly, elliptically, and circularly polarized mid-IR pulses.
(a--c) Time evolution of the phonon amplitudes, $Q_c$, of the nonlinearly coupled $B_2$ modes at 9.05, 19.8, and 25.7~THz for the different polarized excitations of the $E$ mode. The different curves are offset by 0.1~\AA$\sqrt{u}$ for visibility.
(d--f) Time evolution of the ferroelectric polarization, $\langle P_z \rangle$, arising from a rectification of the three $B_2$ modes, and magnetization, $M_z$, arising from the phonon magnetic moment of the $E$ mode. We plot two different orientations and helicities of the polarization of the laser pulse that reverse the direction of the induced ferroelectric polarization ($\pm 45^\circ$) and magnetization (left- and right-handed helicity).
}
}
\label{fig:results}
\end{figure*}


\subsection{Dynamical multiferroic polarization}
%
We now compute the phonon-induced ferroelectric polarization and magnetization for a resonant excitation of the $E$ mode at 26.2~THz, which has the largest mode effective charge of all $E$ modes in the system and hence the largest coupling to the mid-IR pulse. We numerically solve the equations of motion, Eqs.~\eqref{eq:Q_1_time}--\eqref{eq:Q_c}, from which we extract the time-dependent phonon amplitudes of the $E$ mode and the three $B_2$ modes at 9.05, 19.8, and 25.7~THz. The total polarization is a sum of the three $B_2$-mode contributions, $P_z=\sum_{\alpha\in B_2}Z_{\alpha,z} Q_\alpha/V_c$, and the moving average yields the induced ferroelectric polarization, $\langle P_z \rangle$. The magnetization in turn arises directly from the $E$ mode.

In Fig.~\ref{fig:results}, we present the results of the dynamical simulations for a mid-IR pulse with $\omega_0=26.2$~THz, $E_0=25$~MV/cm and $\tau=0.2$~ps. Figs.~\ref{fig:results}(a--c) display the time evolutions of the phonon amplitudes of the three $B_2$ modes for linear, elliptical, and circular excitations of the $E$ mode. For all three excitations, an oscillatory response of the $B_2$ modes is visible. For linear and elliptical excitations, the oscillation is around a quasistatic offset, corresponding to a rectification of the atomic displacements from their equilibrium positions that decays quadratically with the amplitude of the $E$-mode components. In all calculations, the maximum mean-squared atomic displacements well below 10\%{} of the interatomic distance, in accordance with the Lindemann stability criterion \cite{Lindemann1910}. The amplitudes of the $E$-mode components are shown in the Appendix.

Figs.~\ref{fig:results}(d--f) display the time evolutions of the ferroelectric polarization induced by the $B_2$ modes, $\langle P_z \rangle$, as well as the magnetization induced by the $E$ mode, $M_z$. For a linear excitation, the magnetization is zero, $M_z=0$, whereas the ferroelectric polarization is maximal, reaching $\langle P_z \rangle = 2$~$\mu$C/cm$^2$, which is comparable to typical ferroelectric perovskite oxides \cite{khomskii2009classifying}. In contrast, for a circular excitation, the ferroelectric polarization is zero, $\langle P_z \rangle = 0$, whereas the magnetization is maximal, reaching $M_z=0.3$~$\mu_N/V_c$, comparable to previous predictions \cite{Juraschek2019,Geilhufe2021}. For an elliptical excitation, both ferroelectric polarization and magnetization are nonzero and reduced by a factor of $\sqrt{2}$ with respect to their maximum values. An elliptical excitation therefore transiently generates multiferroic polarization in the nonpolar, nonmagnetic material. By changing the orientation or handedness of the polarization of the pulse, both the ferroelectric polarization and magnetization are reversed. Intriguingly, the elliptical excitation leads to an unequal coupling of the two $E$-mode components to the $B_2$ modes, resulting in a splitting of their frequencies (see Appendix) and therefore to an overshoot of the magnetization upon decay. This means that the magnetization changes in the presence of the ferroelectric polarization.


\section{Discussion}
%
We performed calculations for the example of $\gamma$-LiBO$_2$, however the symmetry requirements for the dynamical generation of multiferroic polarization presented here are general to all semiconductors and insulators in the $\bar{4}$ and $\bar{4}2m$ point groups, such as BPO$_4$ \cite{Radaelli2018}. The calculated ferroelectric polarization reaches the magnitude of typical ferroelectrics, whereas the magnetization calculated in the ionic point-charge current picture is rather small. Recent studies have shown however that this picture underestimates the phonon magnetic moment by up to four orders of magnitude due to the neglect of electron-phonon coupling \cite{Cheng2020,Baydin2022,Hernandez2023,Basini2024,Zhang2023_BLG,Geilhufe2023,Merlin2023}. We therefore expect that the possible induced magnetization can be much larger than predicted here, and our calculations should be seen as a lower boundary. We further compute unprecedented strengths of the nonlinear phonon couplings and single-mode anharmonicities that pose the question of whether a perturbative treatment of the phonon dynamics is still valid. This and questions addressing the origin of the large nonlinearities call for comprehensive first-principles and experimental studies in the future. 

Finally, we propose possible experiments to confirm the mechanism presented here. The induced ferroelectric polarization could be measured with time-resolved SHG, whereas the induced magnetization could be measured with time-resolved Faraday or magnetooptic Kerr effect (MOKE) experiments. Both are achievable with state-of-the-art tabletop mid-IR pump/visible probe setups.


\begin{acknowledgments}
We thank C. Romao, D. Bustamante Lopez, and T. Kahana for useful discussions. We acknowledge support from Tel Aviv University. Calculations were performed on local HPC infrastructure. This work was supported by the Israel Science Foundation (ISF) Grant No. 1077/23.
\end{acknowledgments}


\section*{Appendix: Derivations of the nonlinear susceptibilities}


\subsection*{Piezoelectric properties of $\gamma$-LiBO$_2$}

We will revisit the general theory of phononic rectification by nonlinearly driven phonon modes and how this phenomenon can induce a transient ferroelectric polarization in nonpolar materials. First, however, a comment about the piezoelectricity of $\gamma$-LiBO$_2$. Ref.~\cite{Radaelli2018} suggested that $\gamma$-LiBO$_2$, due to its piezoelectric response, possesses the right symmetry requirements for light-induced ferroelectricity. The piezoelectric tensor relates the electric displacement in a crystal, and hence polarization, to an applied stress, which in this case is caused by the ultrafast laser pulse that couples to an optical phonon mode. Ref.~\cite{Radaelli2018} argues that an emergent ferroelectric polarization along the $c$ axis of the crystal requires nonzero piezoelectric tensor components in the third row. The piezoelectric tensor for point group $\bar{4}2 m$ has the form

\begin{equation}
   D_{\bar{4}2 m} =  \begin{pmatrix}
0 & 0 & 0 & d_{14} & 0 & 0\\
0 & 0 & 0 & 0 & d_{14} & 0\\
0 & 0 & 0 & 0 & 0 & d_{36}
    \end{pmatrix}.
\end{equation}
Since $d_{36}$ is nonzero, a ferroelectric polarization may emerge along the $c$-axis from a pump in the $ab$-plane.


\subsection*{Derivation of the equations of motion}

Our model consists of the two components of a degenerate IR-active $E$ mode, denoted as $Q_a$ and $Q_b$, that are driven resonantly by a mid-IR pulse in the $ab$-plane of the crystal, along with a third mode with $B_2$ symmetry, $Q_{c}$, that nonlinearly couples to them through three-phonon coupling. The potential energy associated with these phonon modes can be expressed as 
\begin{equation}
\label{eq:phononpotential_APPENDIX}
    V_{ph}  =  \frac{\Omega_{0}^2}{2}Q_{a}^2+\frac{\Omega_{0}^2}{2}Q_{b}^2+\frac{\Omega_{c}^2}{2}Q_{c}^2 + c Q_{a}Q_{b}Q_{c} + \tilde{V},
\end{equation}
where $\Omega_a=\Omega_b\equiv\Omega_0$ is the eigenfrequency of the $E$ mode and $\Omega_c$ that of the $B_2$ mode. The coefficient $c$ denotes the trilinear, cubic-order, phonon coupling primarily of interest here. The expansion can be continued with

\begin{align}
\tilde{V} = &~ d_a Q_a^4 + d_b Q_b^4 + d_c Q_c^4 \nonumber \\
& + d_{ab} Q_a^2 Q_b^2 + d_{ac} Q_a^2 Q_c^2 + d_{bc} Q_b^2 Q_c^2 \nonumber\\ 
& + g_a Q_a^6 + g_b Q_b^6 + g_c Q_c^6,
\end{align}
which contains the quartic- and sixth-order single-mode anharmonicities and quartic-order nonlinear phonon coupling coefficients, denoted by $d_\alpha$, $g_\alpha$, and $d_{\alpha\beta}$ respectively. Cubic-order single-mode anharmonicities of the type $Q_\alpha^3$, as well as quadratic-linear coupling terms of the type $Q_{a/b}^2 Q_c$ are symmetry-forbidden.

The light-matter interaction of the $E$ modes with the ultrashort mid-IR pulse can be written as
\begin{equation}
    V_{l-m} = -\mathbf{p}_{a/b}\cdot\mathbf{E}(t)
    \label{eq:l-m_appendix} ,
\end{equation}
where $\mathbf{p} = \mathbf{Z}_{a/b}Q_{a/b}$ is the electric dipole moment of the driven phonon modes. $\mathbf{Z}_{a/b} = \sum_{n} \mathbf{Z}^{*}_{n}\mathbf{q}_{n,a/b}/\sqrt{M_n}$ is the mode effective charge vector with $\mathbf{Z}^{*}_{n}$ being the Born effective charge tensor, $\mathbf{q}_{n,a/b}$ the phonon eigenvector, and $M_n$ the atomic mass of atom $n$. The sum runs over all atoms in the unit cell. Since we consider the electric field component of light to be polarized in the $ab$-plane of the crystal, the light-matter coupling of the $B_2$ mode vanishes. The electric field component of the pulse in the material, $\mathbf{E}(t)$, is given by:

\begin{equation}
\mathbf{E}(t)=\mathcal{E}(t)(\cos(\omega_0 t), \cos(\omega_0 t + \phi), 0),
\end{equation}
where $\mathcal{E}(t) = \epsilon_\infty^{-1} E_{0} e^{-t^2/(\tau\sqrt{8\ln(2)})^2}$ is the carrier envelope, $\epsilon_\infty$ is the optical dielectric constant of the material, $E_{0}$ is the peak electric field in free space, $\tau$ is the full width at half maximum pulse duration and $\omega_{0}$ is the center frequency. $\phi$ is the carrier envelope phase that here determines the polarization (linear, elliptical, circular) of the laser pulse. 

The time evolution of the coupled phonon modes can be described by a phenomenological oscillator model derived from the damped Euler-Lagrange equations:

\begin{equation}
    \frac{\text{d}}{\text{d}t}\frac{\partial \mathcal{L}}{\partial \dot{Q}} - \frac{\partial \mathcal{L}}{\partial Q} = -\frac{\partial G}{\partial \dot{Q}}. \label{eq:phononeom_APPENDIX}
\end{equation}
Here, $\mathcal{L}=T - V$ is the phonon Lagrangian containing the phonon kinetic energy, $T=\dot{Q}_a^2/2+\dot{Q}_b^2/2+\dot{Q}_c^2/2$, and the total phonon-dependent potential, $V=V_{ph}+V_{l-m}$,

\begin{align}
V = & ~\frac{\Omega_{0}^2}{2}Q_{a}^2+\frac{\Omega_{0}^2}{2}Q_{b}^2+\frac{\Omega_{c}^2}{2}Q_{c}^2 + c Q_{a}Q_{b}Q_{c} + \tilde{V} \nonumber\\
&-Q_{a}Z_{a,x}E_{x}(t)-Q_{b}Z_{b,y}E_{y}(t).
\end{align}
$G=\kappa_0\dot{Q}_a^2/2+\kappa_0\dot{Q}_b^2/2+\kappa_c\dot{Q}_c^2/2$ is the Rayleigh dissipation function and $\kappa_\alpha$, where $\alpha \in \{a,b,c\}$, is the phonon linewidth, with $\kappa_a=\kappa_b\equiv\kappa_0$. From Eq.~\eqref{eq:phononeom_APPENDIX} we obtain the three coupled equations

\begin{align}
\ddot{Q}_{a} + \kappa_{0}\Dot{Q}_{a} + \Omega^2_{0}Q_{a} & = Z_{a,x}E_{x}(t) - c Q_{b} Q_{c} - \partial_{Q_a} \tilde{V}, \label{eq:Q_1_Appendix} \\
\ddot{Q}_{b} + \kappa_{0}\Dot{Q}_{b} + \Omega^2_{0}Q_{b} & = Z_{b,y}E_{y}(t) - c Q_{a} Q_{c} - \partial_{Q_b} \tilde{V}, \label{eq:Q_2_Appendix} \\
\ddot{Q}_{c} + \kappa_{c}\Dot{Q}_{c} + \Omega^2_{c}Q_{c} & = -cQ_{a}Q_{b} - \partial_{Q_c} \tilde{V}, \label{eq:Q_c_Appendix}
\end{align}
These equations illustrate that the electric field component of light acts as the primary driving force for the IR-active $E$ modes, while the $E$ modes themselves, $Q_{a}Q_{b}$, serve as the main driving force of the coupled $B_2$ mode. The higher order nonlinear coupling terms contained in $\tilde{V}$ are a weaker contribution to the driving force.

To solve these equations analytically, we perform a Fourier transform to the frequency domain. The lowest-order results for $Q_{a}$ and $Q_{b}$ are in first order in the electric field, while the expression for $Q_{c}$ is in second order,

\begin{align}
Q_{a} & = Z_{a,i}\frac{E_{i}(\omega)}{\Delta_{0}(\omega)}, ~ Q_{b} = Z_{b,i}\frac{E_{i}(\omega)}{\Delta_{0}(\omega)}, \label{eq:Q_1_result_Appendix}\\
Q_{c} & = -\frac{c}{\Delta_{c}(\omega)}\left(Q_{a}(\omega)\circledast Q_{b}(\omega)\right)\nonumber\\
\label{eq:Q_c_result_Appendix}
&= -\frac{cZ_{a,i}Z_{b,j}}{\Delta_{c}(\omega)}\left(\frac{E_{i}(\omega)}{\Delta_{0}(\omega)}\circledast \frac{E_{j}(\omega)}{\Delta_{0}(\omega)}\right),
\end{align}
where $\Delta_{\alpha}(\omega) = \Omega^{2}_{\alpha}-\omega^2+i\omega \kappa_{\alpha}$, $\alpha\in\{0,c\}$, and $\circledast$ is the convolution operator. We use the Einstein sum convention for indices denoting the spatial coordinates, $i,j,k$.


\begin{figure*}[t]
\centering
\includegraphics[width=0.7\linewidth]{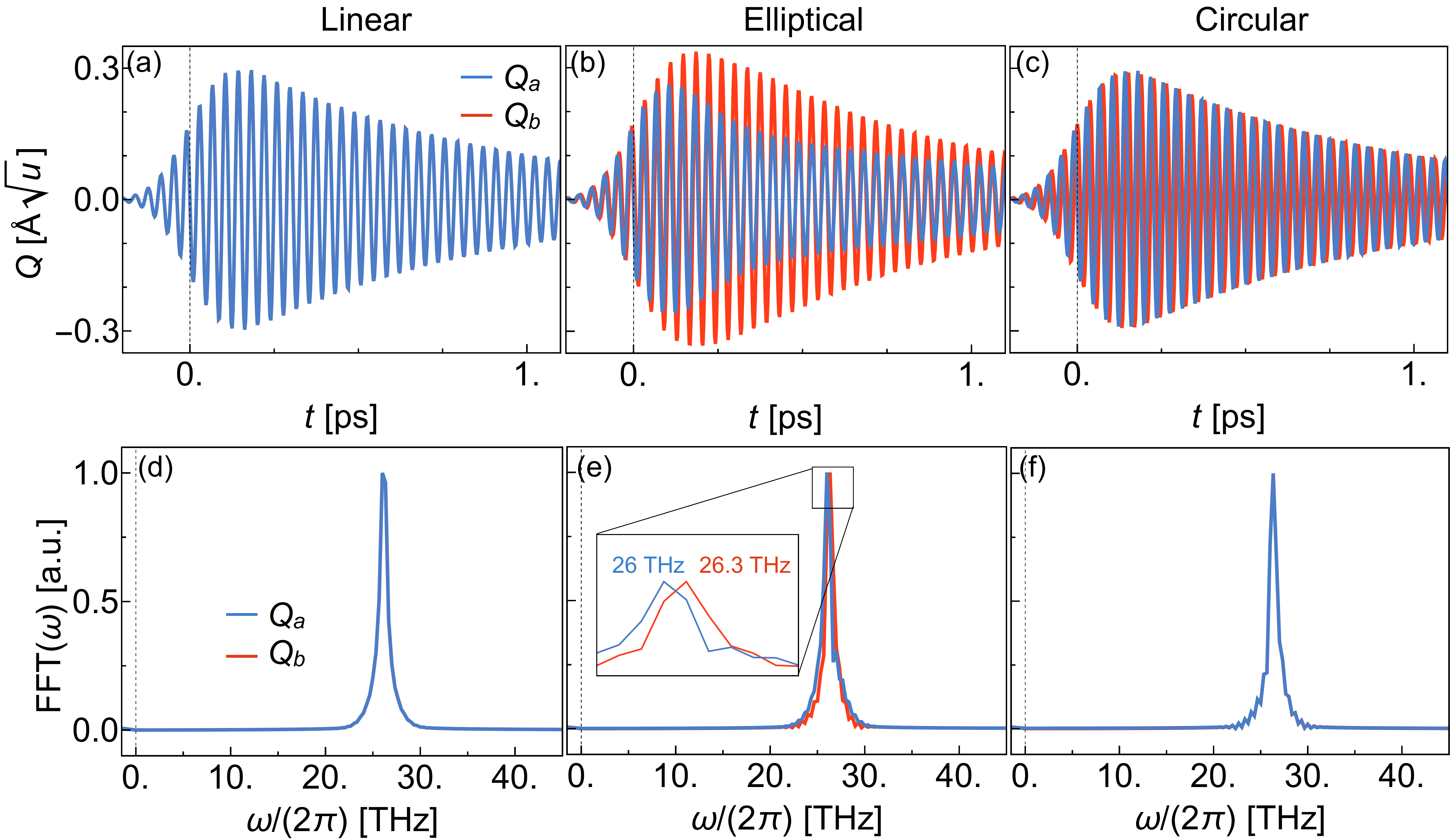}
\caption{
{\small
(a--c) Time-dependent amplitudes of the two orthogonal components of the doubly degenerate $E$(26.2) mode, $Q_a$ and $Q_b$, excited by linearly, elliptically and circularly polarized laser pulses. (d--f) Normalized Fourier transforms of the time traces in (a--c). There is a small splitting in frequency between the two $E$-mode components for the elliptical excitation. 
}}
\label{fig:Emodeamplitudes}
\end{figure*}


\subsection*{Derivation of the nonlinear electric susceptibility}

We now direct our focus towards understanding the induced ferroic polarizations and derive the second-order nonlinear electric susceptibilities discussed in the main text. For the purpose of the derivation, we neglect quartic-order terms in the following, $\tilde{V}=0$. Because the $B_2$ mode produces a polarization along the $c$-axis of the crystal, we can write

\begin{equation}
    P_{c,z}(\omega) = \frac{Z_{c,z}}{V_{c}} Q_{c}(\omega),
\end{equation}
where $V_c$ is the unit-cell volume. Substituting Eq.~\eqref{eq:Q_c_result_Appendix} in the equation above, the analytical expression for the ferroelectric polarization can generally be written in second order of the electric field component of the laser pulse as 

\begin{equation}
P_{c,k}(\omega) = \epsilon_0 \int_{-\infty}^{\infty} \chi^{(2)}_{e,ijk}(\omega,\omega') E_{i}(\omega-\omega')E_{j}(\omega')d\omega' ,
\end{equation}
where $\chi^{(2)}_{e,ijk}$ is the second-order nonlinear electric susceptibility induced by the nonlinear phonon coupling, 

\begin{equation}
\label{eq:electricsusceptibilityAPPENDIX}
\chi^{(2)}_{e,ijk}(\omega,\omega') = \frac{-c}{\sqrt{2\pi}\epsilon_{0}V_c} \frac{Z_{a,i} Z_{b,j} Z_{c,k}}{\Delta_{c}(\omega)\Delta_{0}(\omega-\omega')\Delta_{0}(\omega')}.
\end{equation}
This function characterizes the system's frequency response of the polarization and is particularly strong in the case of resonant driving of the IR-active $E$ modes, where $\omega'=\omega_0=\Omega_{0}$. The ferroelectric polarization corresponds to the static component of the full frequency-dependent polarization response, $P_{c,z}(0)$.


\subsection*{Derivation of the nonlinear magnetoelectric susceptibility}

We now analogously derive the second-order nonlinear magnetoelectric susceptibility. A circular or elliptical superposition of the two IR-active phonon modes $Q_{a}$ and $Q_{b}$ in the $ab$-plane of the crystal produces orbital motions of the ions around their equilibrium positions, generating a magnetization perpendicular to the plane \cite{juraschek2:2017,Juraschek2019,Geilhufe2021,Xiong2022}. The magnetization can be written as

\begin{equation}
    M_{z} = \frac{\mu_{ph}}{V_c}\frac{L_{z}}{\hbar},
\end{equation}
where $\mu_{ph}$ is the phonon magnetic moment and $\mathbf{L} = \mathbf{Q}\times \dot{\mathbf{Q}}$ is the phonon angular momentum with $\mathbf{Q} = (Q_{x},Q_{y},0)$. In the point-charge current picture, $\mu_{ph}=\hbar^{-1}\sum_n Z_n^*/(2M_n)\mathbf{q}_{n,a}\times\mathbf{q}_{n,b}$, where $\mathbf{q}_{n,a/b}$ is the phonon eigenvector of atom $n$, $M_n$ its atomic mass, and the sum runs over all atoms in the unit cell. Substituting the analytical expressions for the phonon amplitudes, $Q_{a}$ and $Q_{b}$, derived according to Eq.~\eqref{eq:Q_1_result_Appendix}, we obtain the magnetization in the frequency domain,

\begin{align}
    M_{k}(\omega) &=\frac{\mu_{ph}}{V_c}\frac{L_{z}(\omega)}{\hbar} \nonumber\\
    &= \frac{\mu_{ph}}{\hbar V_c} \epsilon_{abk} (Q_{a}(\omega)\circledast i\omega Q_{b}(\omega)) \nonumber\\ 
   &= \frac{\mu_{ph} Z_{a,i}Z_{b,j}}{\hbar V_c} \epsilon_{ijk} \left(\frac{E_i(\omega)}{\Delta_0(\omega)}\circledast i\omega \frac{E_j(\omega)}{\Delta_0(\omega)}\right),
\end{align}
where $\epsilon_{ijk}$ is the Levi-Civita tensor. We can cast the above expression into a second-order nonlinear magnetoelectric response of the material to the electric field component of the laser pulse, given by 

\begin{equation}
    M_{k}(\omega) = \int_{-\infty}^{\infty} \chi^{(2)}_{me,ijk}(\omega,\omega')E_{i}(\omega-\omega')E_{j}(\omega') d\omega' ,
\end{equation}
where $\chi^{(2)}_{me,ijk}$ is a second order nonlinear magnetoelectric susceptibility induced by the phonon magnetic moment and given by \cite{BustamanteLopez2024}


\begin{table}[b]
\caption{Calculated IR-active phonon eigenfrequencies in THz for the $B_2$ and $E$ phonons in $\gamma$-LiBO$_2$.}
\begin{tabular}{@{}ll@{\quad}ll@{}}
\hline
Symmetry & $\Omega_c/(2\pi)$ &  \quad Symmetry & $\Omega_0/(2\pi)$ \\ \hline
$B_2$    & \begin{tabular}[t]{@{}l@{}}9.05\\ 19.8\\ 25.7\end{tabular} & \quad $E$ & \begin{tabular}[t]{@{}l@{}}10.2\\ 14.3\\ 15.2\\ 22.1\\ 26.2\\ 28.7\end{tabular} \\
\hline
\end{tabular}
\label{table:phonon_eigen}
\end{table}

\begin{equation}
\label{eq:magnetoelectricsusceptibilityAPPENDIX}
    \chi^{(2)}_{me,ijk}(\omega,\omega') = \epsilon_{ijk} \frac{\mu_{ph}}{\sqrt{2\pi}\hbar V_c} \frac{i(2\omega'-\omega) Z_{a,i}Z_{b,j}}{2\Delta_{0}(\omega-\omega')\Delta_{0}(\omega')} .
\end{equation}
This function characterizes the system's frequency response of the magnetization to the excitation by the laser pulse and is particularly strong in the case of resonant driving of the IR-active $E$ modes, where $\omega'=\omega_0=\Omega_{0}$.


\subsection*{$E$-mode amplitudes}

In Figs.~\ref{fig:Emodeamplitudes}(a--c), we show the time evolution of the amplitudes of the two orthogonal components, $Q_a$ and $Q_b$, of the $E$(26.2) mode. In Figs.~\ref{fig:Emodeamplitudes}(d--f), we show the corresponding Fourier transforms for the time period of $-1.5$~ps to $1.5$~ps, during which the amplitudes are the largest. The elliptical excitation in (e) shows a 0.3~THz splitting of the $E$(26.2) mode, arising from a combination of the trilinear coupling to the three $B_2$ modes, $Q_a Q_b Q_c$, and the quartic-order anharmonicity, $Q_{a/b}^4$. The largest atomic motion induced in the system reaches a maximum of 3\%{} of the interatomic distance, well below the Lindemann stability criterion of about 10\%{}.


\subsection*{Frequency-dependent susceptibilities}

\begin{figure}[t]
\centering
\includegraphics[width=1\linewidth]{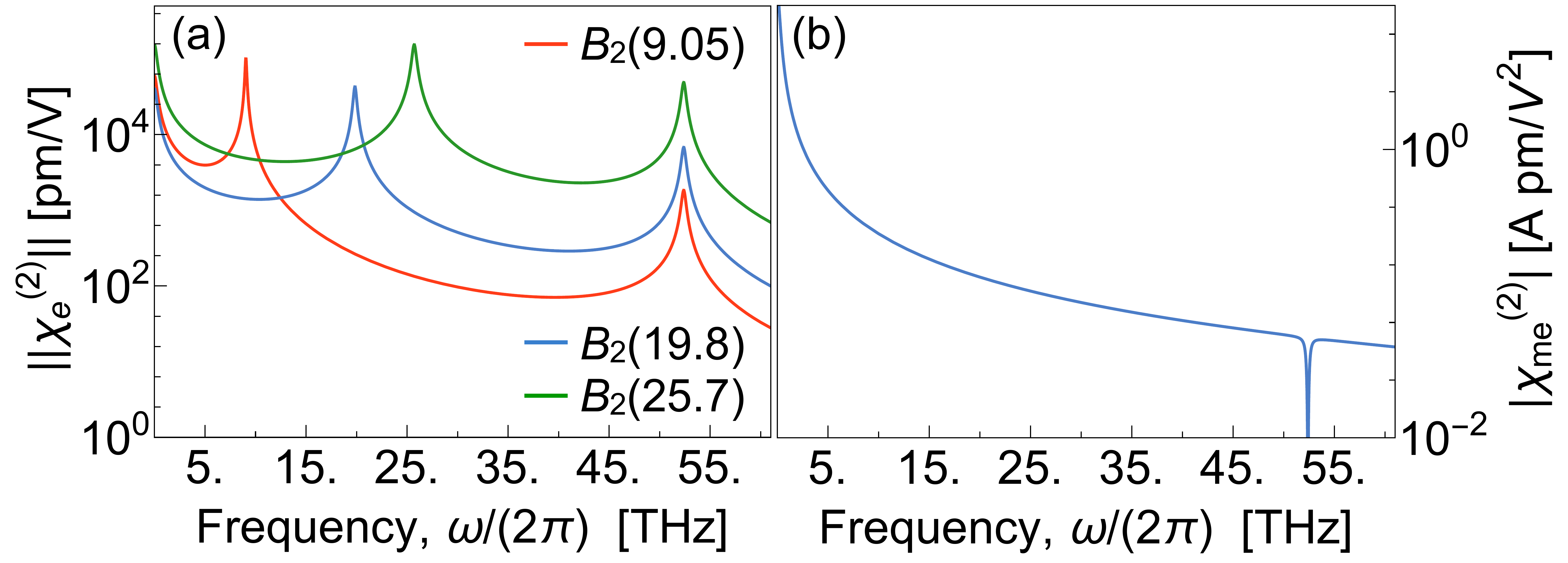}
\caption{
{\small
Phonon-induced susceptibilities.
(a) Norm of the nonlinear electric susceptibility, $||\chi^{(2)}_{e,xyz}(\omega,\omega_0=\Omega_0)||$, arising from the couplings of the $B_2$ modes to the $E$ mode. The three peaks correspond to rectification, oscillatory excitation, and second-harmonic generation (SHG) of the $B_2$ modes. 
(b) Norm of the nonlinear magnetoelectric susceptibility, $||\chi^{(2)}_{me,xyz}(\omega,\omega_0=\Omega_0)||$, arising from the phonon magnetic moment of the $E$ mode. The peak at zero frequency corresponds to the quasistatic component, whereas the dip corresponds to the sum-frequency component.
}
}
\label{fig:susceptibilities}
\end{figure}

In Figs.~\ref{fig:susceptibilities}(a) and (b), we show the norms of the nonlinear electric and magnetoelectric susceptibilities from Eqs.~\eqref{eq:electricsusceptibilityAPPENDIX} and \eqref{eq:magnetoelectricsusceptibilityAPPENDIX}, which characterize the frequency response of the polarization and magnetization. $||\chi_{e}^{(2)}||$ exhibits three peaks corresponding to a rectification, oscillatory excitation, and second-harmonic generation (SHG) of the $B_2$ modes. The SHG component is generally weaker than the former two and negligible in the lattice dynamics. $||\chi_{me}^{(2)}||$ exhibits a peak at zero frequency and a dip at the sum frequency, consistent with the frequency components of the circular driving force, $E_i E_j^* - E_j E_i^*$.


%

\end{document}